# Vortex beams of atoms and molecules

Alon Luski*[1], Yair Segev*†[1], Rea David[1], Ora Bitton[1], Hila Nadler[1], A. Ronny Barnea[2], Alexey Gorlach[3], Ori Cheshnovsky[2], Ido Kaminer[3] and Edvardas Narevicius‡[1]

[1] *Faculty of Chemistry, Weizmann Institute of Science, 7610001 Rehovot, Israel*
[2] *School of Chemistry, Sackler Faculty of Exact Sciences, Tel Aviv University, 6997801 Tel Aviv, Israel*
[3] *Department of Electrical Engineering, Technion - Israel Institute of Technology, 32000 Haifa, Israel*

**Abstract**

Angular momentum plays a central role in a multitude of phenomena in quantum mechanics, recurring in every length scale from the microscopic interactions of light and matter to the macroscopic behavior of superfluids. Vortex beams, carrying intrinsic orbital angular momentum (OAM), are now regularly generated with elementary particles such as photons and electrons, and harnessed for numerous applications including microscopy and communication. Untapped possibilities remain hidden in vortices of non-elementary particles, as their composite structure can lead to coupling of OAM with internal degrees of freedom. However, thus far, the creation of a vortex beam of a non-elementary particle has never been demonstrated experimentally. We present the first vortex beams of atoms and molecules, formed by diffracting supersonic beams of helium atoms and dimers, respectively, off binary masks made from transmission gratings. By achieving large particle coherence lengths and nanometric grating features, we observe a series of vortex rings corresponding to different OAM states in the accumulated images of particles impacting a detector. This method is general and can be applied to most atomic and molecular gases. Our results may open new frontiers in atomic physics, utilizing the additional degree of freedom of OAM to probe collisions and alter fundamental interactions.

**Introduction**

Vortices represent one of the clearest manifestations of angular momentum in nature, governing the hydrodynamics of the smallest insects, the patterns of weather, and the motions of the stars in the galaxy. On all scales, vortices are characterized by the circulation of flux around an axis, whether the flux is comprised of fluid mass in a classical eddy[1], magnetic field in a superconductor, or phase in a superfluid wavefunction[2]. In quantum mechanics, a particle's angular momentum must have an integer value[3]. Accordingly, the quantum manifestations of angular momentum are inherently different from the classical manifestations, contributing to selection rules in optical/electronic transitions, magnetic resonance spectroscopy, and the origins of the atomic fine structure. The twisting of a particle wavefront in a vortex creates an additional intrinsic quantity for that particle: its orbital angular momentum (OAM).

* Equally contributing authors.
†Current affiliation: Department of Physics, University of California, Berkeley
‡ Corresponding author: edn@weizmann.ac.il

The quantization axis of the OAM in a vortex often coincides with the propagation axis of a beam. Such vortex beams were first realized in experiments with photons[4], which have since found applications ranging from optical tweezers[5] to astrophysics[6], communication[7] and more[8,9]. Electron vortex beams[10], first predicted and demonstrated two decades later[11], have been used, for example, to study chirality[12], magnetization mapping[13], and transfer of angular momentum to nanoparticles[14], and ongoing efforts include altering selection rules and controlling nuclear decay[15–17]. Going beyond elementary particles such as the photon and electron, various proposals have been put forth for creating vortex beams of composite particles, such as neutrons[18], protons[19] and atoms[20,21], which are envisioned to alter the fundamental interactions of such particles and to enable probing their internal structure. However, thus far, no experiment has created vortex beams of non-elementary particles[i].

Here we present the first experimental demonstration of vortex states with non-elementary particles, namely atoms and molecules of helium. Starting with a supersonic gas source, we create a beam of atoms with large coherence lengths and diffract them off a nanometric forked grating. The diffracted image reveals a series of "doughnut"-shaped patterns corresponding to non-zero OAM states of helium atoms. We further observe vortex states of helium dimers, a molecule formed in the supersonic expansion of the atomic gas. Our method can be generally applied to any supersonic beam of atoms or molecules, and may be applicable to create vortices of ions and protons.

Several previous demonstrations of vortex beams of photons and electrons utilized spiral phase plates, which create a single OAM state with a topological charge $l$ corresponding to the number of $2\pi$ phases accumulated around the plate's center and equivalent to an OAM of $l\hbar$[22]. However, fabricating such plates becomes challenging for beams of heavy particles with correspondingly shorter wavelengths, such as atoms and molecules, because of the required precision of the spiraling thickness profile. This method is further limited for these particles due to their low transmission probabilities through solid plates. We follow an alternative approach, implementing transmission gratings designed by computer-generated holography, which act as binary amplitude masks to shape the beam so that its far-field pattern acquires the phase of a vortex[23]. These transmission gratings contain edge dislocations that appear as a "fork" shape, where the number of dislocations within the impacting particle's coherence length determines the quanta of OAM imparted to the beam. We demonstrate the production of atomic vortex beams by gratings with one and two dislocations. We also demonstrate molecular vortices produced by a single-edge dislocation grating optimized for the helium dimers.

---

[i] Efforts with neutron beams[42] have been hindered by insufficient coherence lengths[43].

**Experimental setup**

We begin by generating a supersonic beam of helium using a pulsed Even-Lavie valve[24]. This source produces dense ensembles of particles with narrow velocity distributions. The longitudinal wavelength of the particles can be controlled by changing the valve's temperature, where a lower temperature corresponds to a slower beam and thus a longer wavelength. We cool the valve to 115 K to produce a beam with a mean velocity of 1090 m/s, corresponding to a de-Broglie wavelength of 90 pm for helium atoms. At this source temperature, and with a source pressure of about 20 bar, we measure a longitudinal velocity spread of about 3% (FWHM).

The pulse of atoms propagates until it diffracts off the transmission grating. Producing such diffraction gratings for beams of atoms or molecules requires fabrication of nanometric features, as the dimensions must correspond to the relatively small transverse coherence lengths associated with these particles[25–29]. This coherence length is controlled by selecting the range of radial velocities of the particles reaching the grating. We limit the radial spread in our experiment by placing two perpendicular slit skimmers 400 mm downstream of the valve, allowing only a collimated beam to pass through the 150 μm apertures, as depicted in figure 1a. The spread of the impacting particles is further constrained by the solid angle covered by the grating as viewed from the skimmer apertures. We distance the grating 1400 mm from the skimmers to limit the divergence angle to 140 μrad, corresponding to a transverse coherence length of ~700 nm for the beam of helium atoms.

Transmission gratings with forked shapes can produce OAM states with quantized topological charges of $l = \pm n$, where the integer $n$ corresponds to the number of edge dislocations in the hologram. The structure of such a grating can be calculated from the interference of a slightly tilted reference plane wave with a wavefunction containing OAM as the target wave[23]. In practice, such gratings are binary amplitude masks, with areas that either block or transmit the particles. The resulting wavefunction contains multiple quantized OAM states with topological charges $\pm n, \pm 2n, \ldots$ , at corresponding diffraction angles, on either side of a central beam with no angular momentum. The phase singularity in the center of each OAM state leads to a "doughnut" shape in the probability density of the wavefunction. A sample calculated image is depicted in figure 1c.

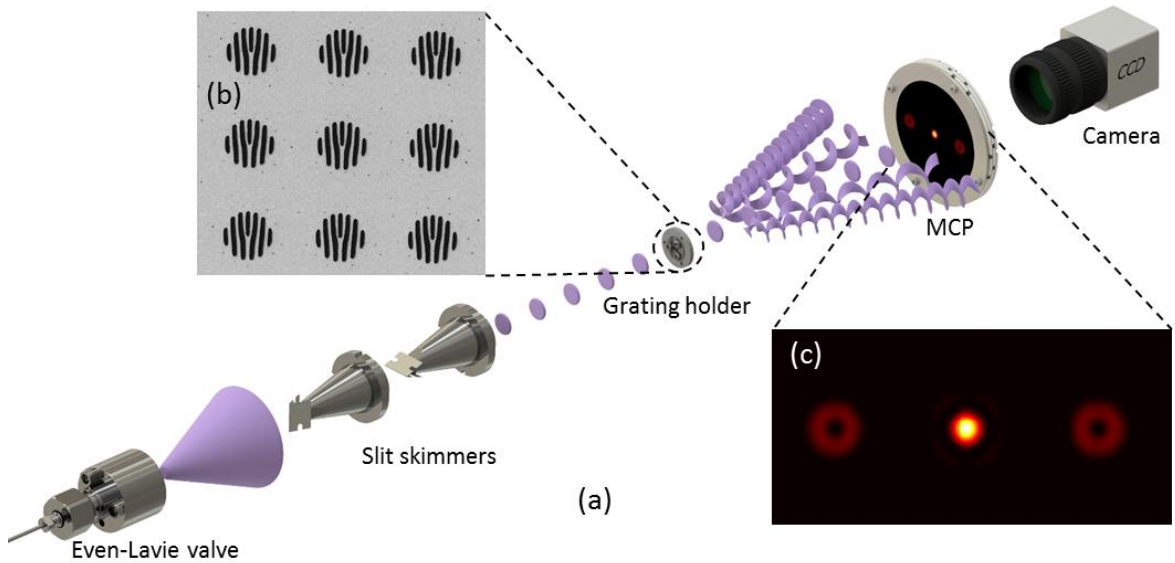

**Figure 1:** Experimental setup for the production and detection of atomic and molecular vortex beams. (a) A cooled Even-Lavie valve emits supersonic pulses of gas (purple cone), which are collimated by two slit skimmers to select the radial velocity spread and increase the transverse coherence length of the particles (purple disks). The beam impacts nanometric gratings, manufactured as an array of forked shapes (b), and diffracted as vortex beams with different values of OAM (purple helixes). The metastable fraction of particles excited by the DBD in the valve decay upon interacting with the MCP, emitting light from a phosphor screen which is recorded on a camera. The impact point is localized in real-time to high resolution, and the integrated image, assembled from hundreds of thousands of events, reveals the characteristic "doughnut" pattern of OAM beams on either side of the central spot (c).

We constructed nanometric binary transmission holograms on 20 nm thick silicon nitride membranes. The fabrication process is described in detail in the supplementary material. Each hologram was etched in a circular area 600 nm in diameter, and the grating slits were designed with a periodicity of 100 nm and approximately equal open and blocked area, to give a diffraction angle of 0.9 mrad for the $l = \pm 1$ orders. To increase the total intensity, the gratings repeat periodically every 1.2 μm over an area of 50x50 μm (figure 1b). The developed area is sufficiently small to avoid incoherent blurring of the image in the far field, while the distance between holograms was chosen to be large enough to avoid cross-talk between neighboring holograms.

For detection, we use a dielectric-barrier discharge[30] (DBD) device installed in the valve, which applies high voltage pulses to produce two different long-lived (metastable) excited states[31]: $2^1S$ and $2^3S$. These states have excess energy of about 20 eV, so virtually any interaction of these atoms with a surface causes relaxation and decay to the ground state. As a result, these metastable states have an almost unity probability for detection by a micro-channel plate (MCP) detector, which we install 1250 mm downstream of the grating. The MCP is assembled with a phosphor screen, and the light emitted from each impact is recorded through a viewport by a camera outside the vacuum chamber. The image on the camera screen is processed in real time to localize each particle's impact position to the resolution of a single detector channel. Note that the DBD excites only a small fraction of the atoms, and therefore this method leaves most of the particles in the vortex beams undetected. The narrow collimation apertures and the low probability of excitation lead to a recorded event rate on the order of one in every fifty valve pulses. The repetition rate of the experiment was set at 30 to 50 Hz, limited by pumping speed in the source chamber and image processing.

**Results**

Figure 2 presents accumulated images of metastable particles impacting the detector. In each image, about 200,000 events were recorded to reveal the far-field pattern corresponding to the probability density of the diffracted matter-wave. The first row of the figure, *a-c*, presents diffraction from a forked grating with a single edge dislocation. The diffraction image contains two clear rings that correspond to atomic vortex beams with topological charges of $l = \pm 1$ on either side of the zero-order beam. The hole in each ring is a result of the phase singularity at the center of each vortex. Higher order OAM states, up to $l = \pm 4$, can be seen in the logarithmic scale image, *c*. The second line of the figure, *d-e*, presents diffraction from a grating with a double edge dislocation. Here, the first two rings correspond to vortex beams with topological charges $l = \pm 2$, and are therefore larger than the first-order rings of the single-edge grating. Higher even-order OAM states, up to $l = \pm 6$, are observable in the logarithmic scale image, *f*.

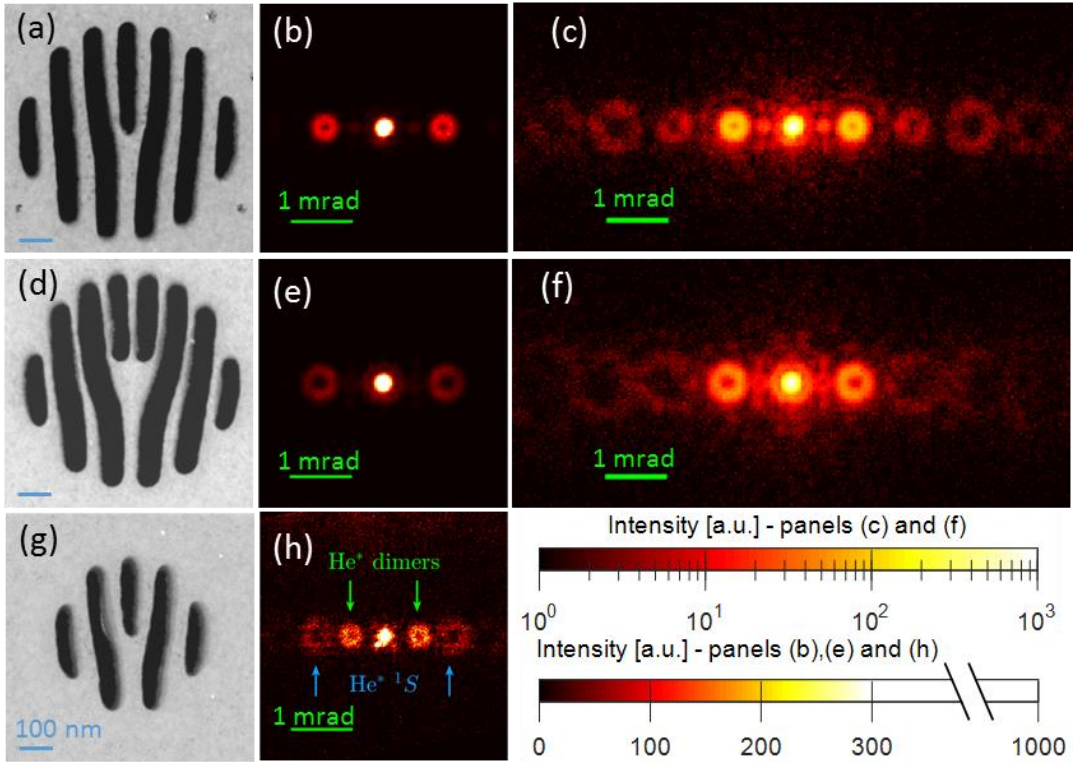

**Figure 2:** Gratings and diffraction images of helium vortex beams, each assembled from 200,000 detected events. (a) A single-edge dislocation grating with a 600 nm diameter produces clear rings (b) corresponding to atomic vortices with OAM of $\pm 1\hbar$ on either side of the central $l = 0$ spot. (c) A logarithmic scale reveals orders up to $l = \pm 4$. (d) A double-edge dislocation grating of the same size produces larger rings (e) with OAM of $\pm 2\hbar$, and the logarithmic scale (f) reveals the $l = \pm 4$ and $l = \pm 6$ orders. The logarithmic scale images (c) and (f) also include "half-order" spots resulting from metastable helium dimers. (g) A smaller, 400 nm grating, optimized for the shorter coherence lengths of these dimers reveals rings (h) associated with OAM-carrying dimer beams. To image the dimer vortices, the detector was distanced further from the gratings, and a 1083 nm laser was used to deflect metastable atoms in the $2^3$S state. The remaining metastable atoms in the $2^1$S state appear as weak rings at twice the dimer $l = \pm 1$ diffraction angles and are convoluted due to coherent cross-talk between neighboring gratings. The color scale in all images is in arbitrary units. The color scale in panels (b), (e) and (h) is saturated above 30% of the maximum intensity to highlight the structure of the rings. The intensity in each pixel is evaluated by counting particle impacts over an area of about 900 micro-radians squared in panels (b), (c), (e) and (f) and 400 micro-radians squared in panel (h).

The diffraction patterns with helium beams reveal additional "half-order" spots, located between the integer diffraction orders expected for atoms. We attribute these spots to metastable helium dimers ($He_2^*$) produced in the valve and excited by the DBD, which have a corresponding de-Broglie wavelength half that of the atoms. Such dimers are routinely created in supersonic valves at similar source conditions[32], especially at low temperatures and high pressures. We verify that these "half-orders" are dimers through the strong dependence of their relative intensity on these source parameters and by observing that a laser locked to the $2^3$S-$2^3$P atomic transition wavelength of 1083 nm[33] only transfers momentum to particles in the integer-order beams. The latter verification is detailed in the supplementary material.

The last row in figure 2, *g-h*, presents the diffraction image from smaller single-edge dislocation gratings, with diameters of 400 nm. These holograms are more suitable for the shorter coherence length of the helium dimer. To compensate for the reduced open area due to the smaller diameter, the gratings in this array repeat every 0.8 μm. Here we use narrower slit skimmers with 75 μm apertures to observe the ring centers by improving the collimation of the beam. We expand the rings of the molecular vortex beam by increasing the free-flight distance from the grating to the detector to 2150 mm. Reducing the grating diameter also contributes to the larger angular expansion of each vortex. To further clarify the image, we deflect the triplet-state metastable atoms with a resonant 1083 nm laser. The singlet state of the metastable atoms remains in the image, and its corresponding rings are convoluted due to cross-talk between the gratings at the decreased spacing.

Figure 3 presents a comparison of the theory to the experimental results. Specifically, we fit the calculated vortex of helium atoms to the intensity profile of the diffracted peaks in the case of the single-edge gratings. The intensity distribution is shown along a line crossing in the transverse image plane through the centers of the rings of different orders. The most significant discrepancy arises from the odd-order peaks of diffracted dimers, absent in the theoretical curve. For the other peaks, we improve the fit by considering that the van der Waals interaction between the grating surface and the atoms narrows the effective width of the slits. Indeed, studies of the interactions between atoms and surfaces made from similar materials showed that van der Waals forces extend to several nanometers away from the surfaces, and reduce the effective width of slits[34]. The effect is enhanced for species with higher polarizability, and is especially strong in beams of metastable helium and neon[35]. In our results, we find an optimal fit for an effective open (transmitting) slit width of 40 nm in place of the 55 nm estimated from the scanning transmission electron microscope (STEM) image. The fit is further improved by adding the contributions of blurring due to pattern repetition of the holograms and of chromatic aberration from the measured beam velocity spread.

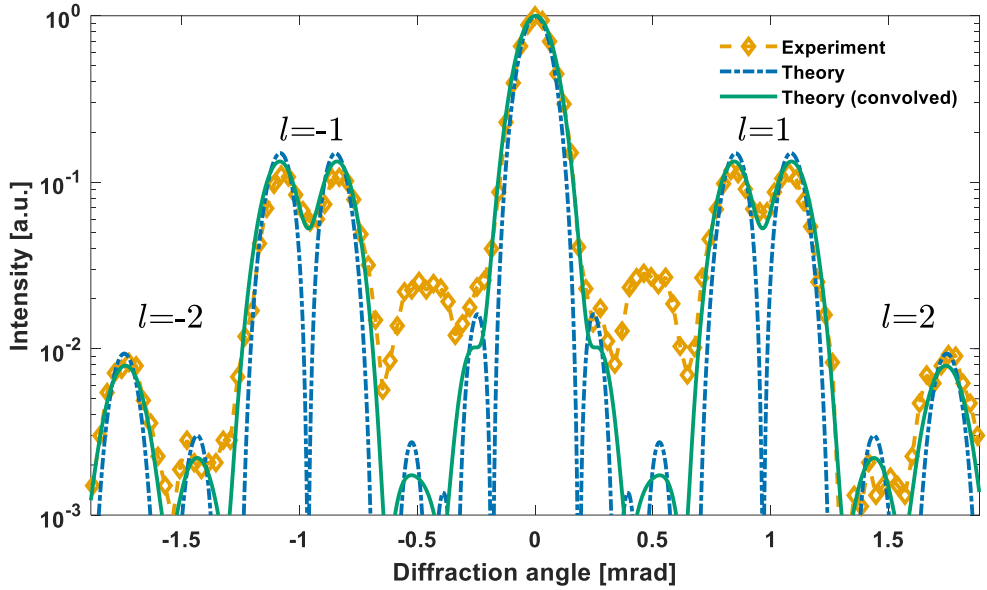

**Figure 3:** Theoretical and experimentally measured intensity curves along a line crossing the helium vortex centers, shown in the range covering the $l = -2$ to $l = 2$ orders. Each point in the experimentally measured curve (orange) represents integration over an area 30 micro-radians wide (along the center-line) by 90 micro-radians vertically. The ideal theoretical curve (blue), calculated for a monochromatic incoming wave of atoms diffracted from a single grating with an open to blocked slit periodicity ratio of 40/60, presents zero intensity at the center of non-zero OAM vortices. The green curve represents the ideal case with the convolution effects due to the velocity spread and the repeating grating pattern in the experiment. The latter curve shows a good fit to the experimentally measured data except for the lack of "half-order" peaks related to helium dimers. All values are given in arbitrary units normalized to the maximum intensity.

**Outlook**

We have demonstrated the generation of vortex beams of atoms and molecules with intrinsic orbital angular momentum. To our knowledge, this is the first successful demonstration of vortices of non-elementary particles. The experiment was enabled by the production of a relatively large coherence length, comparable to the dimensions of the nanometric grating. The method can be applied to most gaseous species due to the generality of the supersonic valve apparatus. Ionizing the atomic beam to create charged vortices could provide an extension to applications of electron vortex beams. A particularly interesting case would be to create OAM beams of protons by ionizing atomic hydrogen, a prospective path to studying their internal structure[19].

The fabrication method we established can be used to design other holograms for wavefront shaping of matter-waves. Shaped matter-waves have a broad range of potential applications; for example, particles may be shaped in such a manner that they will pull a collision target instead of pushing it[36]. In scattering experiments, wavefront shaping opens an entirely new frontier. To date, in all realizations with well-defined low collision energy[31,37,38], the scattering partners are well described by single momentum values, i.e., plane waves. There is often a certain variance of momenta value (transversely and longitudinally), but it is considered as an incoherent distribution rather than a coherent superposition as in our case. OAM beams would introduce chirality to these otherwise axisymmetric probes, and may allow one to explore chiral interactions with various molecular partners.

Conversely, the vortex-generating mask itself may be used as a probe for OAM in scattering experiments. Here, particles exiting a collision are diffracted from a grating, and any pre-existing OAM will lead to an asymmetric intensity image, as demonstrated in OAM sorters of electron beams[39]. In this manner, it would be possible to observe hitherto unexplored properties of reactions or other collisions between atoms or molecules. This concept of identifying OAM resulting from scattering experiments can be generalized beyond atomic physics: OAM sorters can be used in high energy physics to extract new types of information from scattering processes, about chirality and generally regarding the role of transverse wavefunctions in altering fundamental interactions[40].

Finally, the OAM of matter-waves may be used to try to modify selection rules in light-matter interactions. The probability of absorbing light through an atomic transition is dependent on the angular momentum of the photon and on the total angular momentum of the atomic state before and after the transition. Previous works have shown how creating vortex beams of photons can alter selection rules in transitions of trapped ions, with the intrinsic OAM of the photons contributing to the total angular momentum balance[41]. We propose shaping the massive particles as vortices to examine the coupling strength of the matter-wave OAM with other degrees of freedom. Experimental signatures of such an experiment would be, for example, the observation of a transition which would otherwise be forbidden. This approach could be expanded by exciting the atoms to Rydberg states, where the vortex extent and the internal dimensions of the electron wavefunctions in the atoms are comparable. Taking these ideas to light-matter interactions of molecules could couple the OAM degree of freedom to the complexity of molecular rotations, creating novel hybrid rotational states that combine the quantum states of the nuclei and the center of mass.

**Supplementary material**

Grating fabrication

We use SiN membranes, 20 nm thick, on transmission electron microscope grids purchased from TEMwindows. To obtain good adhesion of the electron-beam sensitive resist to the membranes, we apply delicate oxygen and argon plasma in a plasma ashing system, using 150 W plasma power for 5 minutes, and flow rates of 3 sccm for oxygen and 1 sccm for argon at a pressure of 150 mTorr. We spin-coat the membranes with a ZEP-520A resist diluted in anisole (1:1) at a speed of 4000 rpm, resulting in a 100 nm thick resist.

We expose the pattern by electron beam lithography using a RAITH eLINE Plus system. We use a 30 keV accelerating voltage and a 10 microns aperture, yielding a beam current of 25pA. In the exposure step, we apply a dose of 150 µC/cm$^2$. The sample is then developed in n-Amyl acetate for 1 minute and rinsed in isopropyl alcohol for half a minute.

To etch the SiN film, we apply a dry etching technique using the inductively coupled plasma mode in the SPTS Technologies Ltd Rapier system. We apply 50 sccm of $CHF_3$ and 10 sccm of Argon plasma with a coil power of 75W, a bias power of 50 W and a pressure of 20mTorr for 35 seconds. Finally, we add a short oxygen plasma process using the Direct Sample Insertion mode at 100 sccm and a pressure of 35mTorr for 25 seconds, to remove the residual ZEP resist.

Dimer identification

The diffraction patterns with helium beams reveal additional spots centered at half the diffraction angles expected for atoms ("half-orders"). We attribute these spots to helium dimers ($He_2$) produced in the valve and excited to metastable states by the DBD, which have a corresponding de-Broglie wavelength half that of the atoms. Such dimers are routinely created in supersonic valves at similar source conditions[32], especially at low temperatures and high pressures. We verify that these "half-orders" are likely dimers through the strong temperature dependence of their intensity and their lack of response to a laser locked to the $2^3S$-$2^3P$ atomic transition wavelength of 1083 nm. The latter verification is detailed in figure 4 using a diffraction grating made of straight slits with a 100 nm periodicity, which exhibits both the atomic orders and these "half-orders". When we turn on the laser, the particles of atomic helium are deflected and heated, while the "half-orders" remain unaffected. Note that this wavelength does not affect metastable atomic helium in the $2^1S$ state, thus these particles overlap with the even orders of the dimers.

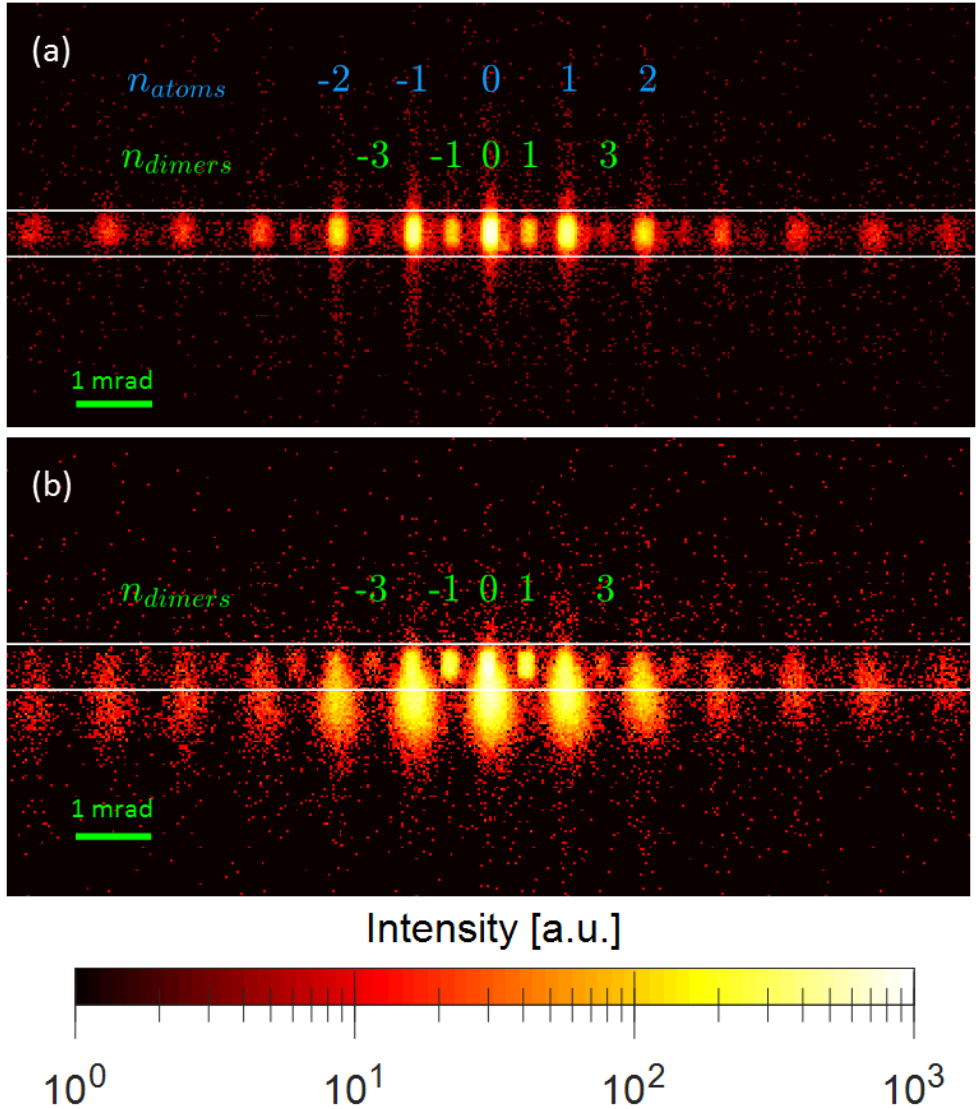

**Figure 4:** Diffraction images of helium beams impacting a regular transmission grating with a slit periodicity of 100 nm. The main diffraction orders are labeled by $n$ and are associated to either atoms or dimers. The atomic beam in (a) freely propagates between the grating and the detector. In (b), a 1083 nm laser, resonant with the $2^3S$-$2^3P$ atomic transition coming from the direction above the detector, intersects the beam after the grating and deflects the triplet-state metastable atoms by photon scattering. The distorted orders belonging to these atoms show broadening due to heating, while the unaffected orders are associated with the helium dimers.